\documentstyle[aps,twocolumn]{revtex}

\begin{document}

\title{Nonequilibrium Self-Interacting Quantum Fields
in Cosmology: The Liouville-Neumann Approach\footnote{Talk presented
at the 5th Thermal Field Theories and Their Applications in Regensburg,
Germany, August 10-14, 1998.}}

\author{Sang Pyo Kim}

\address{Department of Physics, Kunsan National University,
         Kunsan 573-701, Korea}

\date{\today}

\maketitle

\begin{abstract}

We present the so-called Liouville-Neumann (LN) approach
to nonequilibrium quantum fields. The LN approach unifies
the functional Schr\"{o}dinger equation and the LN equation
for time-independent or time-dependent quantum systems
and for equilibrium or nonequilibrium quantum systems.
The LN approach is nonperturbative in that at the lowest
order of coupling constant it gives the same results as those of
the Gaussian effective potential at the zero and finite
temperature in a Minkowski spacetime. We study a self-interacting
quantum field in an expanding Friedmann-Robertson-Walker Universe.
By studying a toy model of anharmonic oscillator and finding
the underlying algebraic structure we propose a
scheme to go beyond the Gaussian approximation.

\end{abstract}


\narrowtext

\section{Introduction}

Recently there have been much interests in the cosmological
scenarios. These models have been developed to
describe the evolution of the Universe
from the early stage to the present epoch \cite{linde}.
Quantum effects or high temperature thermal effects
of matter fields have dominated in the early Universe.
But in the cosmological scenarios the Universe has been
evolving and expanding. Quantum fields in such an expanding
background spacetime could not be described properly as thermal
equilibrium. They are characterized by {\it nonequilibrium
quantum processes}. Quantum nature of the matter fields
predominated as it dated back into the early stage of
the Universe. Several field theoretical methods have been
introduced to describe properly the nonequilibrium quantum process.
Typical and frequently used methods are
closed time path  \cite{ctp}
and thermal field dynamics \cite{tfd}.
Closed time paths integrate over contours of imaginary
(or complex) time and real time. Thermal properties of
quantum field are exhibited through the contours of
imaginary time. Thermal field dynamics doubles
the degrees of freedom of the system and thermal
states are condensation of particles of the system
and its fictitious counterpart.
Despite wide applications closed time paths are
difficult to do integrals for self-interacting quantum fields
and to incorporate various initial quantum conditions.
It would be more than just an interest to have another method,
particularly a canonical method, for nonequilibrium quantum fields.

In this paper we present another field theoretical method,
the so-called {\it Liouville-Neumann approach}, to describe the
nonequilibrium quantum processes in cosmology \cite{kim,bak}.
It is a {\it nonperturbative canonical method}
that unifies the Liouville-Neumann (LN) equation
and the functional Schr\"{o}dinger equation \cite{jackiw}.
It is nonperturbative in that at the lowest
order of coupling constant the wave functionals
take into account nonperturbative quantum effects and
are equivalent to an infinite summation of daisy and
superdaisy diagrams or the Gaussian effective potential.
In particular, it is the LN equation that describes
the nonequilibrium quantum process. The underlying assumption
of the LN approach is that all nonequilibrium processes
are consequence of microscopic processes which are well
described by quantum theory. An introduction of canonical method
will enlarge our understanding of the nonequilibrium  quantum
processes that have mostly been dealt with the closed time
path integral methods but are still not completely
understood. Furthermore, the LN approach has several
advantageous points compared with the path integral methods.
First of all, it is truly canonical. Many useful concepts
and ideas from quantum mechanics can straightforwardly
be applied to quantum field theory by treating a quantum
field as a system with infinite degrees of freedom.
Secondly, the Hilbert space of the exact or approximate
wave functionals is constructed explicitly. Therefore,
it is relatively easy to incorporate various initial
quantum conditions and to evolve them. Moreover, the LN
approach provides a scheme to go beyond the Gaussian approximation.
Thirdly, the LN approach can readily be unified with
thermal field dynamics due to the canonical nature.
Besides these there are also some other useful points.

The organization of this paper is as follows.
In Sec. II we introduce the LN equation to solve
time-dependent quantum systems and unify it with
the functional Schr\"{o}dinger equation to describe
quantum fields. In Sec. III the LN approach is applied
to a massive scalar field in an expanding
Friedmann-Robertson-Walker (FRW) Universe and the quantum states
or Wightmann functions are found for various initial conditions.
In Sec. IV we apply the LN approach to a self-interacting
quantum field in the FRW Universe and find various quantum states
or wave functionals. A toy model of anharmonic oscillator is studied
to go beyond the Gaussian approximation.
Comparison of the LN approach is made with the Gaussian effective
potential.
In Sec. V the LN approach is further exploited
beyond the Gaussian approximation and reveals
the algebraic structure of the anharmonic oscillator.

\section{Liouville-Neumann Approach}

As mentioned in the Introduction
there are several motivations to study the LN
approach to quantum field theory. The LN approach
can be used in a unified way to both time-independent and
time-dependent quantum systems, and to both equilibrium and nonequilibrium
systems. It is a canonical method that deals with the
nonperturbative quantum effects. It is canonical because
it provides a method to solve the functional Schr\"{o}dinger
equation. As a field theoretical method we wish to
unify the LN equation and the functional
Schr\"{o}dinger-picture to make the so-called LN approach.

\subsection{Liouville-Neumann equation}

The LN approach to quantum mechanics
has been introduced by Lewis and Riesenfeld \cite{lewis}
to solve time-dependent quantum systems, especially time-dependent
harmonic oscillators. It is based on the observation that
the Schr\"{o}dinger equation ($\hbar$ = 1)
\begin{equation}
i \frac{\partial}{\partial t}\Psi(t) = \hat{H} (t) \Psi (t)
\end{equation}
for time-dependent as well as time-independent systems
can be solved in  terms of the following operator
\begin{eqnarray}
i \frac{\partial}{\partial t} \hat{{\cal O}}
+ [ \hat{{\cal O}}, \hat{H}]
- i \frac{\dot{\eta}}{\eta} \hat{{\cal O}} = 0,
\nonumber\\
\hat{{\cal O}} \vert \eta_n (t) \rangle = \eta_{n} (t)
\vert \eta_n (t) \rangle.
\end{eqnarray}
The exact quantum states are indeed given by
\begin{equation}
\Psi (t) = \sum_{n} c_n \exp \Bigl[i \int dt \langle
\eta_n (t) \vert i \frac{\partial}{\partial t} - H \vert
\eta_n (t) \rangle \Bigr] \vert \eta_n (t) \rangle.
\end{equation}
In particular, when $\hat{{\cal O}}$ satisfies
the LN equation
\begin{equation}
i \frac{\partial}{\partial t} \hat{{\cal O}}
+ [ \hat{{\cal O}}, \hat{H} ] = 0,
\end{equation}
its eigenvalues are time-independent, $\dot{\eta}_n = 0$.
Such an operator is called the Lewis-Riesenfeld invariant
operator or generalized invariant operator or LN operator.

The aim of this paper is to develop this LN approach
to quantum mechanical systems as a field theoretical
method in the functional Schr\"{o}dinger-picture.

\subsection{Functional Schr\"{o}dinger-Picture}

The functional Schr\"{o}dinger-picture approach
to quantum field theory is a
canonical method in which the wave functionals of a field
as quantum states are determined by the
functional Schr\"{o}dinger equation
\cite{jackiw}
\begin{equation}
i \frac{\partial}{\partial t} \Psi( \phi, t) =
\hat{H} (\phi, - i \frac{\delta}{\delta \phi}) \Psi(\phi, t),
\label{sch eq}
\end{equation}
where $\phi$ represents the quantum field. The Hamiltonian
describes an infinite dimensional system
since the field has infinite degrees of freedom.
The wave functionals $\Psi (\phi, t)$
are specified by assigning $c$-numbers to the field $\phi ({\rm x})$
at a fixed time.
The wave functionals constitute the Hilbert space
equipped with the inner product
\begin{equation}
\langle \Psi_1 \vert \Psi_2 \rangle = \int {\cal D} [\phi]
\Psi_1^* (\phi) \Psi_2 (\phi).
\end{equation}
Operators act on the wave functional of the Hilbert space
\begin{equation}
\hat{{\cal O}} (\phi, \pi) \vert \Psi (\phi, \pi) \rangle
\rightarrow \hat{{\cal O}} ( \phi, - i \frac{\delta}{\delta \phi})
\Psi (\phi).
\end{equation}
The Gaussian wave functionals are a generalization of the
Gaussian states of a harmonic oscillator:
\begin{eqnarray}
\Psi_0 (\phi) &=& N \exp \Bigl[ - \int_{{\bf x}, {\bf y}}
(\phi ({\bf x}) - \bar{\phi}({\bf x}, t))
\nonumber\\
&& \times
\Bigl(\frac{1}{4 G ({\bf x}, {\bf y}, t)}
- i \Sigma ({\bf x}, {\bf y}, t) \Bigr)
(\phi ({\bf y}) - \bar{\phi}({\bf y}, t))
\nonumber\\
&& + i \int_{\bf x} \bar{\pi} ({\bf x}, t)
(\phi ({\bf x}) - \bar{\phi}({\bf x}, t)) \Bigr],
\end{eqnarray}
where
\begin{eqnarray}
\langle \Psi_0 \vert \hat{\phi} ({\bf x}) \vert \Psi_0
\rangle &=& \bar{\phi} ( {\bf x}, t),
\nonumber\\
\langle \Psi_0 \vert \hat{\pi} ({\bf x})
\vert \Psi_0 \rangle &=& \bar{\pi} ( {\bf x}, t),
\nonumber\\
\langle \Psi_0 \vert \hat{\phi} ({\bf x}) \hat{\phi} ( {\bf y})
\vert \Psi_0 \rangle &=& \bar{\phi}({\bf x}, t) \bar{\phi}
({\bf y}, t) + G ({\bf x}, {\bf y}, t),
\nonumber\\
\langle \Psi_0 \vert \hat{\pi} ({\bf x}) \hat{\pi} ( {\bf y})
\vert \Psi_0 \rangle &=& \bar{\pi}({\bf x}, t) \bar{\pi}
({\bf y}, t) + \Sigma ({\bf x}, {\bf y}, t).
\end{eqnarray}

\section{Free Scalar Field in the FRW Universe}

As an application of the LN approach, we consider a
free massive scalar field in a spatially flat
FRW Universe with the metric
\begin{equation}
ds^2 = - dt^2 + R^2 (t) d{\bf x}^2.
\end{equation}
The scalar field has the Lagrangian
\begin{equation}
L = \int d^3x \frac{R^3}{2} \Bigl[\dot{\phi}^2
- \frac{(\nabla \phi)^2}{R^2}
- (m^2 + \xi {}^{(4)}{\rm R} )\phi^2 \Bigr],
\end{equation}
where $\xi = 0, \frac{1}{6}$ for the minimal and conformal
couplings, respectively, and ${}^{(4)} {\rm R}$
is the scalar curvature. The Hamiltonian has the form
\begin{equation}
H = \int d^3x \Bigl[\frac{\pi^2}{2R^3} + \frac{R}{2}
(\nabla \phi)^2 + \frac{R^3}{2} (m^2 + \xi {}^{(4)}{\rm R} ) \phi^2
\Bigr].
\end{equation}
We quantize the scalar field according to the functional
Sch\"{o}dinger-picture
\begin{equation}
i \frac{\partial}{\partial t} \Psi(\phi)
= \hat{H} ( \phi, -i \frac{\delta}{\delta \phi})
\Psi( \phi).
\label{sch eq3}
\end{equation}

Knowing the Fock space of exact quantum states for
a time-dependent harmonic oscillator, we decompose the Hamiltonian
into a sum of harmonic oscillators. For this purpose,
we decompose the field into Fourier-modes
\begin{equation}
\phi({\bf x}, t) = \frac{1}{(2 \pi)^{3/2}}
\int d^3k \phi_{\bf k} (t) e^{i {\bf k} \cdot {\bf x}},
\end{equation}
and refine the field modes as
\begin{eqnarray}
\phi^{(+)}_{\bf k} &=& \frac{1}{2} \Bigl(\phi_{\bf k}
+ \phi_{- {\bf k}} \Bigr),
\nonumber\\
\phi^{(-)}_{\bf k} &=& \frac{i}{2} \Bigl(\phi_{\bf k}
- \phi_{- {\bf k}} \Bigr).
\end{eqnarray}
Space integrations can easily be done for the field and momentum:
\begin{eqnarray}
\int d^3x \phi^2 ({\bf x}, t) &=& \int d^3 k \phi_{\bf k}
\phi_{- {\bf k}} \equiv \sum_{\alpha} \phi_{\alpha}^2,
\nonumber\\
\int d^3x \pi^2 ({\bf x}, t) &=& \int d^3 k \pi_{\bf k}
\pi_{- {\bf k}} \equiv \sum_{\alpha} \pi_{\alpha}^2,
\end{eqnarray}
where $\alpha$ denotes $(\pm, {\bf k})$.
Then the Hamiltonian takes the form
\begin{equation}
\hat{H} = \sum_{\alpha} \Bigl[\frac{\hat{\pi}^2_{\alpha}}{2 R^3}
+ \frac{R^3 \omega_{\alpha}^2}{2} \hat{\phi}_{\alpha}^2
 \Bigr] \equiv \sum_{\alpha} \hat{H}_{\alpha},
 \label{sum os}
\end{equation}
where
\begin{equation}
\omega_{\alpha}^2 (t) = m^2 + \frac{{\bf k}^2}{R^2}
+ \xi {}^{(4)}{\rm R}.
\end{equation}
That is, the Hamiltonian (\ref{sum os}) is a countably infinite
sum of time-dependent harmonic oscillators.
The functional Schr\"{o}dinger equation (\ref{sch eq3}) becomes
the ordinary Schr\"{o}dinger equation for the infinite system
\begin{equation}
i \frac{\partial}{\partial t} \Psi(\phi_{\alpha})
= \sum_{\alpha} \hat{H}_{\alpha} ( \phi_{\alpha},
- i \frac{\partial}{\partial \phi_{\alpha}})
\Psi(\phi_{\alpha}).
\end{equation}
The wave functional is a product of each wave function
\begin{equation}
\Psi (\phi) = \prod_{\alpha} \Psi( \phi_{\alpha}).
\end{equation}
In the end the right hand side will be expressed in terms of the
field $\phi$ by taking an appropriate inverse Fourier-transform.

In the LN approach, one looks for
the following first order operators
\begin{eqnarray}
\hat{a}_{\alpha} &=& i ( \varphi_{\alpha}^* (t) \hat{\pi}_{\alpha}
- \dot{\varphi}_{\alpha}^* (t) \hat{\phi}_{\alpha} ),
\nonumber\\
\hat{a}_{\alpha}^{\dagger} &=& - i ( \varphi_{\alpha} (t) \hat{\pi}_{\alpha}
- \dot{\varphi}_{\alpha} (t) \hat{\phi}_{\alpha} ).
\end{eqnarray}
And then one imposes the LN equations
\begin{eqnarray}
i \frac{\partial}{\partial t} \hat{a}_{\alpha}
+ \bigl[ \hat{a}_{\alpha}, \hat{H}_{\alpha} \bigr] = 0,
\nonumber\\
 i \frac{\partial}{\partial t} \hat{a}^{\dagger}_{\alpha}
+ \bigl[ \hat{a}^{\dagger}_{\alpha}, \hat{H}_{\alpha} \bigr] = 0.
\label{ln op3}
\end{eqnarray}
Equations (\ref{ln op3}) are satisfied when
each mode satisfies the classical equation of motion
\begin{equation}
\ddot{\varphi}_{\alpha} (t) + 3 \frac{\dot{R}}{R}
\dot{\varphi}_{\alpha} (t) + \omega_{\alpha}^2 \varphi_{\alpha}
(t) = 0.
\end{equation}
One further requires $\hat{a}_{\alpha}$
and $\hat{a}_{\alpha}^{\dagger}$ to form the
basis for the Fock space:
\begin{equation}
\bigl[\hat{a}_{\alpha}, \hat{a}_{\beta}^{\dagger}
\bigr] = \delta_{\alpha, \beta}.
\end{equation}
These commutation relations lead to the boundary conditions
on $\varphi_{\alpha}$
\begin{equation}
R^3 (\dot{\varphi}_{\alpha}^* \varphi_{\alpha}
- \dot{\varphi}_{\alpha} \varphi_{\alpha}^* ) = i.
\end{equation}
Then the Fock space of the number states of each mode is
constructed by
\begin{eqnarray}
\hat{a}_{\alpha} \vert 0_{\alpha}, t \rangle &=& 0,
\nonumber\\
\vert n_{\alpha}, t \rangle &=& \frac{1}{\sqrt{n_{\alpha}!}}
\Bigl(\hat{a}_{\alpha} \Bigr)^{n_{\alpha}} \vert 0_{\alpha}, t \rangle.
\end{eqnarray}
The vacuum state of the field is the product of
vacuum states for each mode
\begin{equation}
\vert 0 , t \rangle = \prod_{\alpha} \vert 0_{\alpha},
t \rangle.
\label{vac st3}
\end{equation}
A few comments are in order. The vacuum state (\ref{vac st3})
has all the symmetries of the Lagrangian.
Knowing the Fock space of exact quantum states,
it is easy to incorporate various initial
quantum conditions.

Usually initial conditions are prepared at a particular time
$t_0$. The vacuum and the number states are denoted by
$\vert 0_{\alpha}, t_0 \rangle$ and $\vert n_{\alpha}, t_0
\rangle$. First, with respect to the initial vacuum state,
one finds the Wightmann function
\begin{eqnarray}
\langle \hat{\phi} ({\bf x}, t)&&
\hat{\phi} ({\bf x}', t') \rangle_{\rm vac.}
\nonumber\\
&=& \frac{1}{(2 \pi)^3} \sum_{\alpha_1, \alpha_2}
e^{i ( {\bf k}_1 \cdot {\bf x} - {\bf k}_2 \cdot {\bf x}')}
\langle \hat{\phi}_{\alpha_1} (t)
\hat{\phi}_{\alpha_2} (t') \rangle_{\rm vac.}
\nonumber\\
&=& \int \frac{d^3k}{(2 \pi)^3} e^{ i {\bf k} \cdot ({\bf x}- {\bf x}')}
\varphi_{\bf k}^* (t) \varphi_{{\bf k}'} (t').
\end{eqnarray}
Second, for an initial thermal state
\begin{equation}
\hat{\rho}_{\alpha} (t_0) = \frac{1}{
{\bf Tr} e^{- \beta \hat{H}_{\alpha} (t_0)}}
e^{ - \beta \hat{H}_{\alpha} (t_0) },
\end{equation}
the Wightmann function is found
\begin{eqnarray}
\langle \hat{\phi} ({\bf x}, t)
&&\hat{\phi} ({\bf x}', t') \rangle_{\rm therm.}
=\langle \hat{\phi} ({\bf x}, t) \hat{\phi}
({\bf x}', t') \rangle_{\rm vac.}
\nonumber\\
&&+ \int \frac{d^3k}{(2 \pi)^3} e^{ i {\bf k} \cdot ({\bf x}- {\bf x}')}
\frac{1}{ e^{\beta \omega_{\alpha} (t_0)} - 1}
\nonumber\\
&&\times \Bigl(\varphi_{\bf k}^* (t) \varphi_{{\bf k}'} (t')
+ \varphi_{\bf k} (t) \varphi_{{\bf k}'}^* (t') \Bigr).
\end{eqnarray}
Finally, let us consider a coherent state described by the density operator
\begin{eqnarray}
\hat{\rho}_{\alpha, II}(t_0) = \frac{1}{{\bf Tr}
\hat{\rho}_{\alpha, II} (t_0)}
\exp \Bigl[&-& \beta (\omega_{\alpha} (t_0) \hat{a}_{\alpha}^{\dagger}
\hat{a}_{\alpha}
\nonumber\\
&+& \gamma_{\alpha} \hat{a}_{\alpha}^{\dagger} +
\gamma^*_{\alpha} \hat{a}_{\alpha}
+ \epsilon_{\alpha}) \Bigr],
\label{den2}
\end{eqnarray}
where $\epsilon_{\alpha} = \frac{\gamma_{\alpha}|^2}{\omega_{\alpha}}
 + \frac{\omega_{\alpha}}{2}$.
Equation (\ref{den2}) indeed describes
a thermal state displaced by
\begin{equation}
\hat{D}_{\alpha} = e^{- \frac{\gamma_{\alpha}}{\omega_{\alpha} (t_0)}
\hat{a}_{\alpha}^{\dagger} + \frac{\gamma_{\alpha}^*}{\omega_{\alpha} (t_0)}
\hat{a}_{\alpha}}.
\end{equation}
The Wightmann function is found
\begin{eqnarray}
\langle \hat{\phi} ({\bf x}, t) &&
\hat{\phi} ({\bf x}', t') \rangle_{\rm coh.}
= \langle \hat{\phi} ({\bf x}, t) \hat{\phi} ({\bf x}', t') \rangle_{\rm therm.}
\nonumber\\
&&+ \int \frac{d^3k}{(2 \pi)^3} \varphi_{{\bf k}, c} (t) \varphi_{{\bf k}', c}^* (t'),
\end{eqnarray}
where
\begin{equation}
\varphi_{{\bf k}, c} = \frac{1}{\omega_{\bf k}}
( \gamma_{\bf k} \varphi_{\bf k}^*
+ \gamma_{\bf k}^* \varphi_{\bf k}  ),
\end{equation}
is a classical field.

\section{Self-Interacting Scalar Fields}

In this section we apply the LN approach
to a self-interacting quantum field.
There have been introduced different methods
to investigate the nonperturbative quantum effects.
The most typical nonperturbative method is the Gaussian
effective potential (variational Gaussian approximation)
approach \cite{chang}. But this method
has been applied to time-independent quantum fields.
The Gaussian wave functionals have also been considered
in Ref. \cite{pi}. The problem of solving the covariant
kernel is not completed.  Compared with these methods,
the LN approach provides us with the
explicit wave functionals in terms of the field equations
satisfied by each mode. The LN approach
has recently been applied to quantum fields in curved
spacetime \cite{kim,bak2}.

\subsection{$(0+1)$-Dimensional Toy Model}

To illustrate how the LN approach works for a self-interacting
quantum field, first we consider a time-dependent anharmonic oscillator.
More than just a quantum mechanical system, the
anharmonic oscillator has been used as an important
toy model to test various field theoretical methods.

Let us consider the anharmonic oscillator
\begin{equation}
\hat{H} = \frac{\hat{p}^2}{2}
+ \frac{\omega^2 (t)}{2} \hat{q}^2
+ \frac{\lambda (t)}{4} \hat{q}^4.
\end{equation}
The mass has been rescaled for convenience.
As in the case of harmonic oscillators in Sec. III,
at the lowest order of the coupling constant
one optimizes the Hilbert space by a Fock space
of the annihilation and creation
operators
\begin{eqnarray}
\hat{a} (t) &=& i ( y^* (t) \hat{p} - \dot{y}^* (t) \hat{q}),
\nonumber\\
\hat{a}^{\dagger} (t) &=& - i ( y (t) \hat{p} - \dot{y} (t) \hat{q}).
\end{eqnarray}
We require these operators to minimize the energy
or the Dirac action in the time-independent case or
time-dependent case, respectively. In the time-dependent
case we can not use the minimization of the energy,
because the energy is not conserved.
In the time-independent case, these two principles
of minimization give the identical result. The resulting
quantum state is the Gaussian wave function.
In the time-dependent case, it has been found that
the LN approach gives the same result as
that of the Dirac action principle at the lowest order
of the coupling constant. However, the LN approach
has several advantageous points. First, it is an algebraic
method for the anharmonic oscillator problem. The underlying group
structure proves very convenient and useful. The underlying
structure of the anharmonic oscillator will be studied in detail
in the next section. Second, the quantum states
can be found nonperturbatively to any desired order of
the coupling constant.

Then the anharmonic oscillator has the oscillator representation
\begin{equation}
\hat{H} = \hat{H}_G + \lambda \hat{H}',
\end{equation}
where
\begin{eqnarray}
\hat{H}_G &=& \Bigl[ \dot{y}^* \dot{y} + \omega^2 y^* y
+ 3 \lambda (y^* y)  \Bigr]\Bigl(\hat{a}^{\dagger} \hat{a} +
\frac{1}{2}\Bigr) - \frac{3 \lambda}{4} (y^* y)^2
\nonumber\\
&& + \frac{1}{2} \Bigl[\dot{y}^{*2} + \omega^2
y^{*2} + 3 \lambda (y^* y) y^{*2} \Bigr] \hat{a}^{\dagger 2}
\nonumber\\
&& + \frac{1}{2} \Bigl[\dot{y}^{2} + \omega^2
y^{2} + 3 \lambda (y^* y) y^{2} \Bigr] \hat{a}^{2},
\\
\hat{H}' &=& \frac{1}{4} \sum_{k = 0}^{4} {4 \choose k}
y^{*(4-k)} y^{k} \hat{a}^{\dagger (4-k)} \hat{a}^{k}.
\end{eqnarray}
At the lowest order of the coupling constant, we use
the approximate LN equations
\begin{eqnarray}
i \frac{\partial}{\partial t} \hat{a}
+ \bigl[\hat{a} , \hat{H}_G \Bigr] = 0,
\nonumber\\
 i \frac{\partial}{\partial t} \hat{a}^{\dagger}
+ \bigl[\hat{a}^{\dagger} , \hat{H}_G \Bigr] = 0.
\label{ln eq5}
\end{eqnarray}
Equations (\ref{ln eq5}) lead to the equation for $y$
\begin{equation}
\ddot{y} (t) + \omega^2 (t)  y + 3 \lambda (y^* (t) y (t)) y (t) = 0.
\end{equation}
The state annihilated by $\hat{a}$ becomes an approximate ground state
\begin{equation}
\hat{a} \vert 0 \rangle_{[0]} = 0,
\end{equation}
and in the coordinate representation is given by
\begin{equation}
\Psi_{[0], 0} (q, t) = \Bigl(\frac{1}{2 \pi y^* y} \Bigr)^{1/4}
\exp \Bigl[i \frac{\dot{y}^*}{2 y^*} q^2 \Bigr].
\label{gauss wav}
\end{equation}
Another advantageous point of the LN approach
is that the excited states can be found easily
\begin{equation}
\Psi_{[0], n} (q, t) = \frac{1}{\sqrt{n!}}
\Bigl(\hat{a}^{\dagger} \Bigr)^n
\Psi_{[0],0} (q, t).
\end{equation}
A few comments are in order. Though approximate, these
quantum states take into account nonperturbative
quantum effects already. In field theory,
the Gaussian state (\ref{gauss wav}) is equivalent to
a summation of daisy and superdaisy diagrams.
For a time-independent harmonic oscillator
the Gaussian state (\ref{gauss wav})
is also known to account for the energy within a few percent
even for the strong coupling limit.

We now wish to find the operators that satisfy the LN
equations to any desired order. By treating $\hat{H}'$ perturbatively
in the full LN equations
\begin{eqnarray}
i \frac{\partial}{\partial t} \hat{A}
+ \bigl[\hat{A} , \hat{H} \bigr] = 0,
\nonumber\\
i \frac{\partial}{\partial t} \hat{A}^{\dagger}
+ \bigl[\hat{A}^{\dagger} , \hat{H} \bigr] = 0,
\label{full ln}
\end{eqnarray}
one expands these operators in a series of $\lambda$
\begin{eqnarray}
\hat{A} &=& \hat{a} + \sum_{n = 1}^{\infty} \lambda^{n}
\hat{B}_{(n)},
\nonumber\\
\hat{A}^{\dagger} &=& \hat{a}^{\dagger} + \sum_{n = 1}^{\infty} \lambda^{n}
\hat{B}^{\dagger}_{(n)}.
\label{full op}
\end{eqnarray}
In the Fock space representation,
$\hat{B}_{(n)}$ are again expanded in $\hat{a}$ and
$\hat{a}^{\dagger}$:
\begin{equation}
\hat{B}_{(n)} = \sum_{r, s} b_{(n)}^{(r,s)}
\hat{a}^{\dagger r} \hat{a}^s.
\end{equation}
By comparing the same powers of $\lambda$ in Eq. (\ref{full
ln}), we find the recursive relations
\begin{equation}
\frac{\partial}{\partial t} \hat{B}_{(n)} = i
\bigl[\hat{B}_{(n)}, \hat{H}_G \bigr]
+ i \bigl[\hat{B}_{(n-1)}, \hat{H}' \bigr].
\end{equation}
The recursive relations can be solved iteratively
\begin{eqnarray}
\hat{B}_{(1)} &=& i \int \bigl[\hat{a} , \hat{H}' \bigr],
\nonumber\\
& \vdots&
\nonumber\\
\hat{B}_{(n)} &=& i^n \int \cdots \int \bigl[ \bigl[ \cdots
\bigl[ \bigl[ \hat{a}, \hat{H}' \bigr] , \hat{H}' \bigr], \cdots, \bigr],
\hat{H}'].
\label{rec rel}
\end{eqnarray}
For instance the first order corrections are found
\begin{eqnarray}
\hat{B}_{(1)} &=& 4 \Bigl( i \int y^{*4}\Bigr) \hat{a}^{\dagger 3}
+ 3 \Bigl(i \int y^{*3} y \Bigr)  \hat{a}^{\dagger 2} \hat{a}
\nonumber\\
&& + 2 \Bigl(i \int y^{*2} y^2 \Bigr) \hat{a}^{\dagger} \hat{a}^{2}
+ \Bigl(i \int y^* y^3 \Bigr) \hat{a}^3.
\end{eqnarray}
An improved ground state is obtained from
\begin{equation}
\hat{A}_{[1]} \vert 0 \rangle_{[1]} = 0,
\end{equation}
where
\begin{equation}
\hat{A}_{[1]} = \hat{a} + \lambda \hat{B}_{(1)}.
\end{equation}
This procedure can be repeated to solve
Eq. (\ref{full ln}). The quantum states
at any order of the coupling constant are obtained therefrom.

A comment is in order. In the language of field theory,
the wave functional determined by $\hat{a}$ and $\hat{a}^{\dagger}$
corresponds to a free propagator with a renormalized mass and that determined by
$\hat{A}$ and $\hat{A}^{\dagger}$ in Eq. (\ref{full op}) corresponds
to a summation of all loop diagrams. In this sense
Eq. (\ref{rec rel}) can be interpreted as the Feynman rules on the
closed time path for the time-dependent system.

\subsection{Self-Interacting Scalar Field}

We now turn to a self-interacting quantum field.
A quantum field is equivalent to a system of infinitely
mode-decomposed harmonic or anharmonic oscillators.
Thus the results from the anharmonic oscillator
system are expected to be useful in studying the
self-interaction quantum field.
Though the formalism (LN approach) to be developed
in this paper can be applied
to any quantum field theory, either renormalizable or
nonrenormalizable, we shall confine our attention to
the $\phi^4$-theory. The reason is that the results from
our formalism can readily be compared with those from
other methods.

The $\phi^4$-theory in the FRW
Universe has the Hamiltonian density
\begin{equation}
H = \frac{\pi^2}{2 R^3}  + \frac{R}{2}
( \nabla \phi)^2 + R^3 \Bigl(\frac{m^2}{2} \phi^2
+ \frac{\lambda}{4} \phi^4 \Bigr).
\end{equation}
Before the Fourier-decomposition,
the field is divided into a classical background field
and quantum fluctuations
\begin{equation}
\phi = \phi_c + \phi_f.
\end{equation}
The classical background field depends only on the
comoving time and equals to the space average of the field
(zero-mode).
Quantum mechanically it is a coherent state of the field
\begin{equation}
\langle \hat{\phi} \rangle = \phi_c (t).
\end{equation}
We assume the fluctuations to have symmetric quantum states
\begin{equation}
\langle \hat{\phi}_f \rangle = 0.
\end{equation}
Then the Hamiltonian density is the sum of
classical background field part and quantum fluctuations
part:
\begin{eqnarray}
H &=& \Biggl[\frac{\pi_c^2}{2 R^3}  + R^3 \Bigl(\frac{m^2}{2}
\phi_c^2 + \frac{\lambda}{4} \phi_c^4 \Bigr) \Biggr]
+ \Biggl[\frac{\pi_f^2}{2 R^3} + \frac{R}{2}
( \nabla \phi_f)^2
\nonumber\\
&&+ R^3 \Bigl(\frac{m^2}{2}
+ \frac{3 \lambda}{2} \phi_c^2 \Bigr) \phi_f^2
+ \frac{\lambda R^3}{4} \phi_f^4 \Bigr) \Biggr]
\nonumber\\
&& + \Biggl[\frac{\pi_c \pi_f}{R^3}
+ R^3 \bigl(m^2 + \lambda \phi_c^3 \bigr) \phi_f +
\lambda R^3 \phi_c \phi_f^3  \Biggr].
\label{full ham}
\end{eqnarray}
The expectation values of the terms in the last square
bracket vanish with respect to the symmetric quantum states
of fluctuations. So these terms will not considered any more.

First, we study the classical background field.
Only those terms whose expectation values do not vanish
with respect to the symmetric quantum states of fluctuations
contribute to the Hamiltonian density for the classical background field
\begin{equation}
H_c = \frac{\pi_c^2}{2 R^3} + R^3 \Bigl(\frac{m^2}{2}
+ \frac{3 \lambda}{2} \langle \hat{\phi}_f^2 \rangle \Bigr)
\phi_c^2 + \frac{\lambda}{4} \phi_c^4.
\end{equation}
The classical background field is frequently treated as a classical system.
Then the equation of motion is
\begin{equation}
\ddot{\phi}_c + 3 \frac{\dot{R}}{R} \dot{\phi}_c
+ (m^2 + 3 \lambda \langle \hat{\phi}_f^2 \rangle ) \phi_c
+ \lambda \phi_c^3 = 0.
\end{equation}
The infinite contribution from the quantum fluctuations
should be regulated in a suitable manner.
But more proper treatment is to quantize even the classical
background field. Therefore, the background field is nothing but
an anharmonic oscillator with the mass shifted to
$m^2 + 3 \lambda \langle \hat{\phi}_f^2 \rangle$.
The classical background field plays the role of an inflaton
in the inflationary scenarios. The importance
of such a quantum background field has been shown in Ref. \cite{bak2}.
The annihilation and creation operators of the Fock space
are
\begin{eqnarray}
\hat{a}_c &=& i ( \varphi_c^* \hat{\pi}_c
- R^3 \dot{\varphi_c}^* \hat{\phi}_c)
- i ( \varphi_c^* {\pi}_c
- R^3 \dot{\varphi_c}^* {\phi}_c) ,
\nonumber\\
\hat{a}_c^{\dagger} &=& - i ( \varphi_c (t)\hat{\pi}_c
- R^3 \dot{\varphi_c} \hat{\phi}_c)
+ i ( \varphi_c \pi_c
- R^3 \dot{\varphi_c} \phi_c).
\label{back an-cr}
\end{eqnarray}
The $\hat{a}_c$ and $\hat{a}_c^{\dagger}$
are chosen so that the vacuum state gives the expectation values
\begin{eqnarray}
\langle \hat{\phi}_c \rangle &=& \phi_c,
\nonumber\\
\langle \hat{\pi}_c \rangle &=& \pi_c =
R^3 \dot{\phi}_c.
\end{eqnarray}
Requiring the LN equations for
$\hat{a}_c$ and $\hat{a}_c^{\dagger}$,
it is shown that $\varphi_c$ satisfies the equation
\begin{equation}
\ddot{\varphi}_c + 3 \frac{\dot{R}}{R} \dot{\varphi}_c
+ (m^2 + 3 \lambda \langle \hat{\phi}_f^2 \rangle ) \varphi_c
+ \lambda (\varphi_c^* \varphi_c) \varphi_c  = 0.
\label{back eq2}
\end{equation}
In the above procedure, $\phi_c$ is required to satisfy the same equation
(\ref{back eq2}) as $\varphi_c$.
So the terms proportional to
the identity operator in Eq. (\ref{back an-cr}) become
the Wronskian of the equation of motion, and constant.
This makes our procedure self-consistent.
In the minimal uncertainty proposal for the
inflation \cite{bak2}, a particular choice
of coherent state was assumed
\begin{equation}
\phi_c = \pi_c = 0,
\end{equation}
and the quantum states of the inflaton
were again the symmetric Gaussian states.

Next, we consider the quantum fluctuations.
The terms in the second square bracket of Eq. (\ref{full ham})
give rise to
the Hamiltonian density for quantum fluctuations
\begin{equation}
H_f = \frac{\pi_f^2}{2 R^3} + \frac{R}{2}
( \nabla \phi_f)^2 + R^3 \Bigl(\frac{m^2}{2}
+ \frac{3 \lambda}{2} \phi_c^2 \Bigr) \phi_f^2
+ \frac{\lambda R^3}{4} \phi_f^4.
\end{equation}
Upon decomposing into Fourier-modes, we keep
only those terms with nonvanishing contribution
\begin{equation}
\int d^3 x \phi_f^4 ({\bf x}, t) =
\frac{3}{(2 \pi)^3} \sum_{\alpha \neq \beta}
\phi^2_{\alpha} \phi^2_{\beta}
+ \frac{1}{(2 \pi)^3} \sum_{\alpha} \phi^4_{\alpha}.
\end{equation}
We express the fluctuations and their momenta
in terms of the annihilation and creation
operators
\begin{eqnarray}
\hat{\phi}_{\alpha} &=& \varphi_{\alpha} \hat{a}_{\alpha}
+ \varphi_{\alpha}^* \hat{a}^{\dagger}_{\alpha},
\nonumber\\
\hat{\pi}_{\alpha} &=& R^3 \Bigl(\dot{\varphi}_{\alpha} \hat{a}_{\alpha}
+ \dot{\varphi}_{\alpha}^* \hat{a}^{\dagger}_{\alpha} \Bigr).
\end{eqnarray}
As for the anharmonic oscillator, we find the oscillator
representation of the Hamiltonian for fluctuations
\begin{equation}
\hat{H}_f = \hat{H}_{f, G} + \lambda \hat{H}'_f,
\end{equation}
where $\hat{H}_{f, G}$ and $\hat{H}'_f$ are the quadratic and non-quadratic
parts. At the lowest order of the coupling constant we are
interested in $\hat{H}_{f, G}$:
\begin{eqnarray}
\hat{H}_{f, G} &=& R^3 \sum_{\alpha} \Biggl[\dot{\varphi}_{\alpha}^*
\dot{\varphi}_{\alpha} + \Bigl( m^2 + \frac{{\bf k}^2}{R^2}
+ 3 \lambda \phi^2_c \Bigr) \varphi_{\alpha}^*
\varphi_{\alpha}
\nonumber\\
&&{} + 3 \lambda \Bigl(\frac{1}{(2 \pi)^3} \sum_{\beta}
\varphi_{\beta}^* \varphi_{\beta} \Bigr) \varphi_{\alpha}^*
\varphi_{\alpha} \Biggr] \Bigl(\hat{a}^{\dagger}_{\alpha}
\hat{a}_{\alpha} + \frac{1}{2} \Bigr)
\nonumber\\
&&{} - \frac{3 \lambda R^3 }{4} \frac{1}{(2 \pi)^3}
\Bigl(\sum_{\alpha} \varphi_{\alpha}^*
\varphi_{\alpha} \Bigr)^2
\nonumber\\
&& + \frac{R^3}{2} \sum_{\alpha} \Biggl[\dot{\varphi}_{\alpha}^{*2}
+ \Bigl( m^2 + \frac{{\bf k}^2}{R^2}
+ 3 \lambda \phi^2_c \Bigr) \varphi_{\alpha}^{*2}
\nonumber\\
&&{} + 3 \lambda \Bigl(\frac{1}{(2 \pi)^3} \sum_{\beta}
\varphi_{\beta}^* \varphi_{\beta} \Bigr) \varphi_{\alpha}^{*2}
\Biggr] \hat{a}^{\dagger 2}_{\alpha}
\nonumber\\
&& + \frac{R^3}{2} \sum_{\alpha} \Biggl[\dot{\varphi}_{\alpha}^{2}
+ \Bigl( m^2 + \frac{{\bf k}^2}{R^2}
+ 3 \lambda \phi^2_c \Bigr) \varphi_{\alpha}^{2}
\nonumber\\
&&{} + 3 \lambda \Bigl(\frac{1}{(2 \pi)^3} \sum_{\beta}
\varphi_{\beta}^* \varphi_{\beta} \Bigr) \varphi_{\alpha}^{2}
\Biggr] \hat{a}^{2}_{\alpha}.
\end{eqnarray}
The equations of motion for the variables $\varphi_{\alpha}$
\begin{eqnarray}
\ddot{\varphi}_{\alpha} &&+ 3 \frac{\dot{R}}{R}
\varphi_{\alpha} + \Bigl( m^2 + \frac{{\bf k}^2}{R^2}
+ 3 \lambda \phi^2_c \Bigr) \varphi_{\alpha}
\nonumber\\
&& + 3 \lambda \Bigl(\frac{1}{(2 \pi)^3} \sum_{\beta}
\varphi_{\beta}^* \varphi_{\beta} \Bigr) \varphi_{\alpha} = 0,
\label{fl eq}
\end{eqnarray}
are determined from the LN equations
\begin{eqnarray}
i \frac{\partial}{\partial t} \hat{a}_{\alpha}
+ \bigl[\hat{a}_{\alpha} , \hat{H}_{f, G} \bigr] = 0,
\nonumber\\
 i \frac{\partial}{\partial t} \hat{a}^{\dagger}_{\alpha}
+ \bigl[\hat{a}^{\dagger}_{\alpha} , \hat{H}_{f, G} \bigr] = 0.
\end{eqnarray}
The state annihilated by all annihilation operators
is an approximate ground state
\begin{equation}
\hat{a}_{\alpha} \vert 0 , t \rangle_{[0]} = 0.
\end{equation}
The vacuum expectation value of the Hamiltonian for
fluctuations is
\begin{eqnarray}
{}_{[0]} \langle 0, t \vert \hat{H}_{f, G} \vert
0 , t \rangle_{[0]} =
\frac{R^3}{2} \sum_{\alpha} \Biggl[\dot{\varphi}_{\alpha}^*
\dot{\varphi}_{\alpha}
+ \Bigl( m^2 + \frac{{\bf k}^2}{R^2}
\nonumber\\
+ 3 \lambda \phi^2_c \Bigr) \varphi_{\alpha}^*
\varphi_{\alpha}
+ \frac{3 \lambda}{2} \Bigl(\frac{1}{(2 \pi)^3} \sum_{\beta}
\varphi_{\beta}^* \varphi_{\beta} \Bigr) \varphi_{\alpha}^*
\varphi_{\alpha} \Biggr].
\end{eqnarray}
The number states are similarly constructed by
\begin{equation}
\vert n_{\alpha}, t \rangle_{[0]} = \frac{1}{\sqrt{n_{\alpha}!}}
\Bigl( \hat{a}^{\dagger}_{\alpha} \Bigr)^{n_{\alpha}}
\vert 0, t \rangle_{[0]}.
\end{equation}
The quantum state of each mode is a linear
superposition of its number states.
The quantum state of the fluctuations is a product
of such quantum states
\begin{equation}
\Psi_{f, [0]} ( \phi_f) = \prod_{\alpha}
\Psi_{\alpha, [0]} (\phi_{\alpha}).
\end{equation}

As emphasized before, the LN approach constructs
the Hilbert space for the classical background field
and fluctuations on which various initial conditions
can be relatively easily incorporated. This LN
approach is expected to be a powerful and convenient
method especially for studying the early Universe
and the time-dependent process of critical phenomena.

\subsection{Comparison with the Gaussian Effective
Potential}

In order to compare with the results from other methods,
let us consider the Minkowski spacetime
\begin{equation}
R (t) = 1.
\end{equation}
The solutions to Eq. (\ref{fl eq}) are easily found
\begin{equation}
\varphi_{\alpha} = \frac{1}{\sqrt{2 \Omega_{\alpha}}}
e^{-i \Omega_{\alpha} t}.
\end{equation}
The $\Omega_{\alpha}$ satisfies the gap equation
\begin{equation}
\Omega^2_{\alpha} = {\bf k}^2 + \mu^2 (\phi_c),
\end{equation}
where
\begin{equation}
\mu^2 (\phi_c) = m^2 + 3 \lambda \phi_c^2
+ 3 \lambda \frac{1}{(2 \pi)^3} \sum_{\beta}
\frac{1}{2 \Omega_{\beta}}
\end{equation}
is the renormalized mass.
From the equations of the motion, one is left with
the quadratic Hamiltonian
\begin{eqnarray}
\hat{H}_{f, G} &=&
R^3 \sum_{\alpha} \Biggl[\dot{\varphi}_{\alpha}^*
\dot{\varphi}_{\alpha} + \Bigl( m^2 + \frac{{\bf k}^2}{R^2}
+ 3 \lambda \phi^2_c \Bigr) \varphi_{\alpha}^*
\varphi_{\alpha}
\nonumber\\
&&{} + 3 \lambda \Bigl(\frac{1}{(2 \pi)^3} \sum_{\beta}
\varphi_{\beta}^* \varphi_{\beta} \Bigr) \varphi_{\alpha}^*
\varphi_{\alpha} \Biggr] \Bigl(\hat{a}^{\dagger}_{\alpha}
\hat{a}_{\alpha} + \frac{1}{2} \Bigr)
\nonumber\\
&&{} - \frac{3 \lambda R^3 }{4} \frac{1}{(2 \pi)^3}
\Bigl(\sum_{\alpha} \varphi_{\alpha}^*
\varphi_{\alpha} \Bigr)^2.
\end{eqnarray}

First, we compute the effective potential at the zero temperature.
The effective potential is obtained by taking the expectation
of the Hamiltonian (\ref{full ham})
except the term $\hat{\pi}_c^2$:
\begin{equation}
V_{\rm eff.} (\phi_c, \mu) =
 \frac{m^2}{2} \phi_c^2
+ \frac{\lambda}{4} \phi_c^4 +
I_1 (\mu) - \frac{3 \lambda}{2} I^2_0 (\mu),
\end{equation}
where
\begin{eqnarray}
I_0 (\mu) &=& \frac{1}{(2 \pi)^3} \sum_{\alpha}
\frac{1}{\Omega_{\alpha}} = \frac{1}{2} \int
\frac{d^3k}{(2\pi)^3} \frac{1}{\Omega_{\bf k}},
\nonumber\\
I_1 (\mu) &=& \frac{1}{(2 \pi)^3} \sum_{\alpha}
\frac{1}{2} \Omega_{\alpha} = \frac{1}{2} \int
\frac{d^3k}{(2\pi)^3} \Omega_{\bf k}.
\end{eqnarray}
We have thus shown that the effective potential
is equal to the Gaussian effective potential \cite{chang}.

Next, we compute the effective potential at finite
temperature. The effective potential equals to the
Helmholtz free energy
\begin{equation}
F = - \frac{1}{\beta} \ln (Z) = V_{\rm eff.},
\end{equation}
where $Z$ is the partition function
\begin{equation}
Z = {\rm Tr} e^{- \beta \hat{H}}.
\end{equation}
To the first order of the coupling constant,
the partition function is given by
\begin{eqnarray}
Z &=& {\rm Tr} e^{- \beta(\hat{H}_{f, G} + \lambda
\hat{H}'_{f})}
\nonumber\\
&\cong& {\rm Tr} e^{- \beta \hat{H}_{f, G}} -  \lambda
\beta {\rm Tr} \Bigl[ \hat{H}'_{f} e^{- \beta \hat{H}_{f, G}}
\Bigr]
\end{eqnarray}
The finite temperature effective potential is found
\begin{equation}
V^T_{\rm eff.} (\phi_c, \mu)  = \frac{m^2}{2} \phi_c^2
+ \frac{\lambda}{4} \phi_c^4
+ (I_1 + I_1^{\beta} ) - \frac{3 \lambda}{2}
(I_0 + I_0^{\beta} )^2,
\end{equation}
where
\begin{eqnarray}
I_0^{\beta} (\mu) &=& \frac{1}{(2 \pi)^3} \sum_{\alpha}
\frac{1}{\Omega_{\alpha}} \frac{1}{e^{\beta \Omega_{\alpha}}
-1}
\nonumber\\
&=& \int
\frac{d^3k}{(2\pi)^3} \frac{1}{\Omega ({\bf k})}
\frac{1}{e^{\beta \Omega (\bf k) } -1},
\nonumber\\
I_1^{\beta} (\mu) &=& \frac{1}{(2 \pi)^3 \beta} \sum_{\alpha}
\ln \Bigl(1 - e^{- \beta \Omega_{\alpha}} \Bigr)
\nonumber\\
&=& \frac{1}{\beta} \int
\frac{d^3k}{(2\pi)^3} \ln \Bigl(1 - e^{- \beta \Omega ({\bf k})} \Bigr).
\end{eqnarray}
The finite temperature effective potential can also be rewritten as
\begin{equation}
V^T_{\rm eff.} (\phi_c, \mu) = V_{\rm eff.} (\phi_c, M)
+ \frac{1}{\beta} \int
\frac{d^3k}{(2\pi)^3} \ln \Bigl(1 -
e^{- \beta \Omega ({\bf k})} \Bigr),
\end{equation}
in terms of the renormalized mass
\begin{equation}
M^2 = m^2 + 3 \lambda (I_0 + I_0^{\beta} + \phi_c^2 ).
\end{equation}
Hence the effective potential at finite temperature
is equal to the Gaussian effective potential at
finite temperature.
We have shown that the LN approach to
quantum field theory at the lowest order of the
coupling constant gives the same results as
those of the Gaussian effective potential approach.
The real vantage of the LN approach is that
one can compute readily the effective potential
beyond the Gaussian effective potential to any desired order
as will be sketched in the next section.

\section{Beyond the Gaussian Approximation}

We introduce an algebraic, nonperturbative method to
quantum field theory.
As a $(0+1)$-dimensional toy model for a self-interacting
quantum field, we shall consider again the anharmonic oscillator
\begin{equation}
\hat{H} = \frac{1}{2} \hat{p}^2 + \frac{m^2}{2} \hat{q}^2
+ \frac{\lambda}{4} \hat{q}^4.
\label{gen ham}
\end{equation}
One of results of the NL approach is that the
anharmonic oscillator has an algebraic structure
which has not been much noticed in the literature.
In the Fock space we can represent the Hamiltonian in the form
\begin{equation}
\hat{H} = \Omega_G \hat{A}^{\dagger} \hat{A} + E_{\rm vac.}.
\end{equation}
The operators $\hat{A}$ and $\hat{A}^{\dagger}$ do not satisfy
the standard commutation relation. They rather satisfy a generalized
deformed algebra of the form
\begin{equation}
\hat{A} \hat{A}^{\dagger} = F(\hat{A}^{\dagger}\hat{A}).
\label{gen q-a}
\end{equation}
The standard commutation relation is given by
$F(x) = x +1$, and the $q$-deformed algebra by
$F(x) = Q^2 x + 1$. In the Fock space representation
$F$ is in general a polynomial of $x$.

It has been found that up to the first order of the coupling
constant the anharmonic oscillator has the
supersymmetric form \cite{bak}
\begin{equation}
\hat{H} = \frac{\Omega_{[1]}}{2} \Bigl(
\hat{A}_{[1]}^{\dagger}\hat{A}_{[1]}
+ \hat{A}_{[1]} \hat{A}_{[1]}^{\dagger}\Bigr),
\end{equation}
where the frequency is determined from the gap equation
\begin{equation}
\Omega_{[1]} = \Omega_G - \frac{3 \lambda}{8 \Omega_G^3}.
\end{equation}
The explicit form of $\hat{A}$ and $\hat{A}^{\dagger}$
is found from the LN equations.
Remarkably these operators satisfy the $q$-deformed algebra
\begin{equation}
\hat{A}_{[1]} \hat{A}_{[1]}^{\dagger} - Q^2
\hat{A}_{[1]}^{\dagger} \hat{A}_{[1]} = 1,
\end{equation}
where
\begin{equation}
Q^2 = 1 + \frac{3 \lambda}{4 \Omega^3_G}.
\end{equation}
The first improved vacuum state beyond the Gaussian approximation
is defined by
\begin{equation}
\hat{A}_{[1]} \vert 0 \rangle_{[1]} = 0.
\end{equation}
The $q$-deformed algebra enables one to find the excited
states
\begin{equation}
\vert n \rangle_{[1]} = \frac{1}{\sqrt{[n]!}}
\Bigl(\hat{A}_{[1]}^{\dagger} \Bigr)^n \vert 0 \rangle_{[1]},
\end{equation}
where
\begin{equation}
[n] = \frac{Q^{2n} - 1}{Q^2 - 1}.
\end{equation}
It is straightforward to obtain the energy eigenvalues
\begin{equation}
E_{[1], n} = \frac{\Omega_{[1]}}{2} \Bigl([n] + [n+1] \Bigr).
\end{equation}

To any order of the coupling constant, we are able to find
the Hamiltonian in the form (\ref{gen ham}). For instance, to
the third order of the coupling constant, we find the
Hamiltonian
\begin{equation}
\hat{H} = \hbar \Omega_G \hat{A}^{\dagger}_{[3]}
\hat{A}_{[3]} + E_{[3], {\rm vac.}},
\end{equation}
where
\begin{eqnarray}
E_{[3], {\rm vac.}} &=& \hbar \Omega_G \Biggl[ \frac{1}{2}
- \frac{3}{16}\Bigl(\frac{\hbar \lambda}{\Omega_G^3} \Bigr)
- \frac{3}{128}\Bigl(\frac{\hbar \lambda}{\Omega_G^3} \Bigr)^2
\nonumber\\
&&{}{} + \frac{25}{1024}\Bigl(\frac{\hbar \lambda}{\Omega_G^3} \Bigr)^3
 \Biggr].
 \end{eqnarray}
$E_{[3], {\rm vac.}}$ is the vacuum energy
up to the same order of the coupling constant.
The generalized deformed algebra (\ref{gen q-a})
is found to be
\begin{eqnarray}
\hat{A}_{[3]}\hat{A}^{\dagger}_{[3]}
&=& 1 + \hat{A}^{\dagger}_{[3]} \hat{A}_{[3]} + \frac{3}{4}
\frac{\hbar \lambda}{\Omega_G^3} \hat{A}^{\dagger}_{[3]} \hat{A}_{[3]}
\nonumber\\
&& - \Bigl(\frac{\hbar \lambda}{\Omega_G^3} \Bigr)^2 \Bigl\{
\frac{9}{32} + \frac{3}{16}\hat{A}^{\dagger}_{[3]} \hat{A}_{[3]}
+ \frac{69}{64} (\hat{A}^{\dagger}_{[3]} \hat{A}_{[3]})^2 \Bigr\}
\nonumber\\
&& + \Bigl(\frac{\hbar \lambda}{\Omega_G^3} \Bigr)^3 \Bigl\{
\frac{387}{512} + \frac{141}{64}\hat{A}^{\dagger}_{[3]} \hat{A}_{[3]}
\nonumber\\
&& + \frac{279}{512} (\hat{A}^{\dagger}_{[3]} \hat{A}_{[3]})^2
+ \frac{633}{256} (\hat{A}^{\dagger}_{[3]} \hat{A}_{[3]})^3
\Bigr\}.
\end{eqnarray}
Therefore, one obtains the better improved vacuum state
beyond the Gaussian approximation from
\begin{equation}
\hat{A}_{[3]} \vert 0 \rangle_{[3]} = 0.
\end{equation}
As in the harmonic oscillator case, the excited states are obtained
by applying  $\hat{A}^{\dagger}$ to the vacuum state:
\begin{equation}
\vert k \rangle_{[3]} = \frac{1}{\sqrt{\prod^{k}_{l =
0} N_k }} \Bigl(\hat{A}_{[3]} \Bigr)^k \vert 0 \rangle_{[3]},
\end{equation}
where
\begin{equation}
\hat{A}^{\dagger}_{[3]} \hat{A}_{[3]} \vert k \rangle_{[3]}
= N_k \vert k \rangle_{[3]}.
\end{equation}
The eigenstates are constructed to be orthonormal
\begin{equation}
{}_{[3]}\langle k \vert l \rangle_{[3]} = \delta_{kl}.
\end{equation}
It is straightforward to see that the eigenvalues satisfy
the following recursive relations
\begin{equation}
N_k = F(N_{k-1}),~ N_0 = 0.
\end{equation}
Therefore, the anharmonic oscillator has the energy
\begin{equation}
E_{[3],k} = \hbar \Omega_{G} N_k + E_{[3], {\rm vac.}}.
\end{equation}

Though not shown explicitly, the algebraic structure
of the quartic anharmonic oscillator with q-deformed algebra
at the first order of the coupling constant or with the
generalized deformed algebra at the higher order of coupling
constant is expected to play
a pivotal role in nonperturbative quantum field theory.
For instance the q-deformed algebra results in
the partition function of the anharmonic oscillator
valid even for the strong coupling limit. The extension of the
q-deformed algebra to the self-interacting quantum field
in Sec. IV will give rise to much more correctly
the effective potential at the zero
or finite temperature than the Gaussian effective potential.

\section{Summary and Outlook}

In this paper we have introduced the so-called
{\it Liuoville-Neumann (LN) approach} to nonequilibrium quantum fields.
It is a canonical method that unifies the functional
Schr\"{o}dinger equation with the LN equation.
The LN approach is nonperturbative in that at the lowest order
of the coupling constant it gives the same results as the Gaussian effective
potential. The LN approach is universal since it is equally
applicable to time-independent and time-dependent quantum systems,
and to equilibrium and nonequilibrium systems.
Though not shown in this paper, the LN approach can be
unified with thermal field dynamics by doubling the degrees
of freedom. This is achieved by introducing a fictitious Hamiltonian
$\tilde{\hat{H}}$ and enlarging the Hilbert space constructed
in this paper. The LN approach combined with thermal field
dynamics is expected to give more physical intuition to
our understanding of nonequilibrium quantum fields.
The renormalization problem and the partition function
up to higher orders of the coupling constant
are not dealt with.
Interestingly enough, fermion systems and gauge theories
are challenging problems to the LN approach.

\acknowledgments

The author would like to thank Dr. D. Bak, Dr. J.-Y. Ji,  Prof. S. K. Kim,
Prof. K.-S. Soh and Prof. J. H. Yee for some parts of this
work and many useful discussions. This work was
supported in parts by the Center for Theoretical Physics,
Seoul National University and by the Non-Directed Research
Fund, Korea Research Foundation, 1997.

\end{document}